\documentclass[a4paper,11pt]{article}
\pdfoutput=1 
\usepackage{jheppub} 
\usepackage{etoolbox}
\patchcmd{\maketitle}{\@fpheader}{}{}{}

\usepackage[T1]{fontenc} 
\usepackage[utf8]{inputenc} 
\usepackage{microtype} 
\usepackage{lmodern} 
\usepackage{mdframed} 
\usepackage{graphicx}
\usepackage{cancel} 
\usepackage{todonotes} 
\usepackage[normalem]{ulem} 
\usepackage{soul} 
\usepackage{amsmath}
\usepackage{amsfonts}
\usepackage{amssymb}
\usepackage{stmaryrd}
\usepackage{mathtools}
\usepackage{mathrsfs}

\usepackage{mleftright} \mleftright
\usepackage{tensor} 
\usepackage{braket} 
\usepackage{mathrsfs} 

\usepackage{bbold}
\usepackage{color}
\usepackage{braket}
\usepackage{slashed}
\usepackage{hyperref}

\newcommand{\comment}[1]{}

\DeclareMathAlphabet{\mathfs}{U}{rsfs}{m}{n}                     %

\newcommand{\be}{\nopagebreak[3]\begin{equation}}
\newcommand{\ee}{\end{equation}}
\newcommand{\bee}{\nopagebreak[3]\begin{equation*}}
\newcommand{\eee}{\end{equation*}}
\newcommand{\ba}{\nopagebreak[3]\begin{eqnarray}}
\newcommand{\ea}{\end{eqnarray}}
\newcommand{\baa}{\nopagebreak[3]\begin{eqnarray*}}
\newcommand{\eaa}{\end{eqnarray*}}

\title{Stringy (Galilei) Newton-Hooke Chern-Simons Gravities}
\author[1,2]{Luis Avil\'es,}
\author[3]{Joaquim Gomis,}
\author[1,2]{Diego Hidalgo}

\affiliation[1]{Centro de Estudios Cient\'ificos (CECs), Av. Arturo Prat 514, Valdivia, Chile,}
\affiliation[2]{Departamento de F\'isica, Universidad de Concepci\'on, Casilla 160-C, Concepci\'on, Chile,}
\affiliation[3]{Departament de F\'isica Qu\`{a}ntica i Astrof\'isica and Institut de Ci\`{e}ncies del Cosmos (ICCUB), Universitat de Barcelona, Mart\'i i Franqu\`{e}s 1, E-08028 Barcelona, Spain}
\emailAdd{aviles@cecs.cl}
\emailAdd{joaquim.gomis@gmail.com}
\emailAdd{dihidalgot@gmail.com}
\preprint{ICCUB-19-008}

\abstract{We construct  Chern-Simons gravities in $(2+1)$-dimensional space-time considering the Stringy Galilei algebra both with
and without non-central extensions. In the first case, there is an invariant and non-degenerate bilinear form, however the field equations do not allow to  express the spin connections in terms of the dreibeins. In the second case there is no invariant non-degenerate bilinear form.
Therefore, in both cases we do not have an ordinary gravity theory.
Instead, if we consider the stringy Newton-Hooke algebra with extensions as gauge group we have an  invariant non-degenerate metric
and from the field equations we express the spin connections in terms of the geometric fields.
 }

\begin{document}

\maketitle
\flushbottom

\section{Introduction}

The renewed interest  to study non-relativistic (NR) gravities is due in part to the possible applications of them to NR holography in condensed matter \cite{Sachdev,Liu}. The construction of NR gravity theories in the target space can be done at least in three different ways: i) as a  NR limit of a suitable relativistic theory \cite{Havas:1964zza,Trautman,Datcour}
, ii) a pure NR construction using as a tool a NR symmetry algebra \cite{DePietri:1994je, DePietri:1994rd,Horava:2009uw,Andringa:2010it}, iii) or a null reduction of a relativistic theory \cite{Duval:1984cj,Julia:1994bs,Bergshoeff:2017dqq,Obers:2017gk}.

There is another possibility to construct gravity from a world sheet point of view. We could consider a 2d non-linear sigma string action,
in a 
gravitational background and compute the corresponding beta functions of the gravitational couplings  \cite{Friedan:1980jm,Callan:1985ia}. 
The vanishing of the beta functions, that implies the conformal invariance at quantum level,  gives the  equations of motion of 
Einstein gravity plus  the $\alpha'$ corrections. 



In particular, the action of a NR string can be obtained as a NR limit of a relativistic string action. The  NR limit is not unique, in the case of a string 
we have two limits \cite{Batlle:2016iel}: 1) the particle limit, where one only scales the target space time coordinate, 2) the string limit, where a target temporal
and space component are scaled. In the first case one obtains a non-vibrating string, which is a collection of massless galilean particles \cite{Sou}.
The Lagrangian is strictly invariant under the ordinary 
Galilei transformations and the associated Noether charges verify the Galilei 
algebra with no central extension. The most general point transformation that leave the Lagrangian invariant is infinite dimensional, 
therefore with no extensions in the algebra \cite{Batlle:2016iel}.
In the second case it is found a vibrating string which is unitary and UV completed \cite{Gomis:2000bd}. The symmetry transformations
close under the stringy Galilei transformations \cite{Brugues:2004an, passerini}, see also \cite{Barducci:2018wuj}.
The algebra of the Noether charges contains two non-central charges: a vector  
$Z_a$ and antisymmetric tensor $Z_{ab}$, with $a,b=0,1$ being the target space longitudinal indexes, i.e., the directions where the
string lives in target space.  Also in this case the algebra of point transformations is infinite dimensional.

In a recent work \cite{Gomis:2019zyu} considering vibrating non-relativistic strings in a general string Newton-Cartan background 
\cite{Bergshoeff:2018yvt}
and computing the associated beta functions it was found the stringy Newton-Cartan equations of motions. In an earlier work these equations were postulated by applying the gauging procedure to the stringy Galilei algebra \cite{Andringa:2012uz}, however, these have not been obtained from an action principle.
On the other hand, it has been recently constructed an extended string Newton-Cartan gravity action in $(3+1)$-dimensions which can be obtained  as the NR limit of the Einstein-Hilbert action coupled to an auxiliary two-form and one-form gauge field \cite{Bergshoeff:2018vfn}. In contrast with the relativistic theory, not all components of the spin connection are completely determined and these occur as Lagrange multipliers in a second order nonrelativistic theory. 
The dimensional reduction of the action to 3d gives the {\it particle} Chern-Simons extended Bargmann gravity introduced in \cite{Papageorgiou:2009zc,Bergshoeff-Rosseel,Obers-Hartong}, and not the string Newton-Cartan gravity in 3d.

In this paper we will study the construction of a $(2+1)$-dimensional stringy Chern-Simons (CS) gravity by considering the stringy Galilei algebra 
\cite{Brugues:2004an, passerini,Andringa:2012uz}
both with and without extensions. For the case of the non-extended stringy Galilei algebra there exits an invariant and non-degenerate bilinear form
and a 3d CS action can be constructed. However, the corresponding equations of motion do not allow to express all the components of the boost spin connections in terms of the dreibeins.

In a second order formulation the undetermined part of the spin connection  part can be interpreted as a Lagrange multiplier for a geometrical constraint\footnote{ An analogous situation occurs in Galilei gravity \cite{Bergshoeff:2017btm}.}.
  On the other hand, if we construct the CS action with the extended stringy Galilei algebra, the associated bilinear form is degenerate.
This property can be seen easily using the double extension formalism \cite{medina,Figueroa-OFarrill:1995opp}, (see also \cite{Matulich:2019}). One of the main issues which appears with a degenerate bilinear form is that some of the curvatures do not vanish and therefore not all dynamics fields can be determined.
The NR limit of a string moving in a AdS background leads to a non-relativistic action that is invariant under the stringy Newton-Hooke symmetries;
 in order to overcome the above mentioned difficulty of the spin connections we consider the stringy Newton-Hooke algebra with extensions\footnote{
This algebra can be also obtained by the double extension procedure that we discuss in the Appendix.} \cite{Brugues:2006yd},

 We consider the CS gravity associated to the stringy Newton-Hooke algebra, this fact allows to express all the boost and Lorentz spin connections in terms of the corresponding geometric fields. The action in this case has the property that the dynamical fields associated with all non-central central extensions  appear explicitly in the action. This property is new with respect to previous discussions \cite{Bergshoeff:2018yvt}.

 This paper is organized as follows. In section 2 we will apply the gauging technique to the Stringy Galilei algebra ($\mathfrak{sga}$) in $(2+1)$-dimensions. In particular, we will find that the expressions of the independent spin connections of this geometry 
 are not completely determined. 
 This fact is in agreement with the fact that it   is not possible to apply the double extension procedure to this algebra.

On the other hand, in section 3 we will gauge the $(2+1)$-stringy Newton-Hooke with extensions algebra which contains 
$Z_a$ and antisymmetric tensor $Z_{ab}$, with $a,b=0,1$ extra generators.
Finally, we give some comments on the second order formulation that can be constructed for the action principle obtained in Section 3.
We have a section 4 where we give  some  conclusions and outlook. In an Appendix we study the double extension procedure applied to our algebras.

\label{sec:introduction}

\section{Stringy Galilei Chern-Simons gravities}

Let us consider the $(2+1)$-stringy Galilei algebra without extensions \cite{Brugues:2004an, passerini,Andringa:2012uz}

\begin{align} \label{SGAlgebra}
[H_a,M_b]=-\eta_{ab} P ,\quad [N,H_a]={\epsilon_a}^b H_b ,\quad [N,M_a]={\epsilon_a}^b M_b,
\end{align}
where $H_a$ are the longitudinal translations, $M_a\equiv M_{2a}$ the boost transformation, $N\equiv M_{01}$ the longitudinal Lorentz transformation, and $P\equiv P_2$ the transverse translation, with $a=0,1$. These generators admit the following invariant bilinear form

\begin{equation}\label{SGPairing}
\langle H_a,M_b\rangle=\gamma_1 \epsilon_{ab}, \quad \langle N,P \rangle=-\gamma_1, \quad \langle N,N \rangle=\gamma_2.
\end{equation} 

We notice that in contrast to Galilei particle case, this bilinear form is non-degenerate\footnote{ We will call the ``trace'', which was denoted by $\big< \cdots \big>$ in Eq.\eqref{SGPairing}, invariant if 
\begin{align}
Z \cdot \langle X, Y \rangle := \langle [Z, X] , Y \rangle + \langle X, [Z,Y] \rangle \equiv 0,
\end{align} 
for all Lie algebra elements $X, Y$ and $Z$, and symmetric and non-degenerate if $\Omega_{AB}= \big< X_A, X_B \big> $ satisfy $\Omega^{AB}\Omega_{BC} =\delta^{A}_{C}$.
 } for all real $\gamma_i$ and therefore we can construct a CS action \cite{Achucarro:1987vz} \cite{Witten:1988}. Let us introduce the gauge fields $\tau^a$ and $e$ associated to the time and spatial translations satisfying the following relations

\begin{equation} \label{relations}
e_\mu e^\nu = \delta^\nu_\mu -{\tau_\mu}^a {\tau^\nu}_a , \quad \tau^{\mu a} \tau_{\mu b} = \delta^a_b ,\quad  {\tau^\mu}_a e_\mu = 0,
\end{equation}
 and the the boost and Lorentz spin connections denoted by $\omega$ and $\omega^a$, respectively. These one-form fields $\tau, e , \omega$ and $\omega_a$ can be combined into the 1-form connection $A$ of the CS theory as
\begin{equation}
A = \tau^a H_a+eP+ \omega^a M_a + \omega N. 
\end{equation}

Then, the CS action yields
\begin{eqnarray}\label{SGaction}
\nonumber S [A ] & = & \int \big< A \wedge dA + \frac{1}{3}  [A,A] \wedge A \big>\\
& =& \int 2\gamma_1(\epsilon_{ab} \tau^a d\omega^b-\omega de +\omega \tau_a \omega^a)+\gamma_2 \omega d \omega.
\end{eqnarray}

The equations of motion obtained from \eqref{SGaction} are
\begin{eqnarray} \label{motionEq1}
\delta_{\tau^a}S: \quad  && \gamma_1 R^a(M)= 0,\\ \label{motionRM1}
\delta_\omega S: \quad && \gamma_2 R(N)-\gamma_1 R(P)=0,\\ \label{motionRN1}
\delta_{\omega^a} S: \quad  && \gamma_1 R^a (H) = 0,\\ \label{motionRH1}
\delta_e S: \quad && \gamma_1 R(N)= 0,
\end{eqnarray}
where,
\begin{eqnarray}
R^a(H)&=& d\tau^a-\epsilon^{ab}\omega \tau_b ,\\
R^a(M)&=& d\omega^a-\epsilon^{ab}\omega \omega_b,\\
R(N)&=& d\omega,\\
R(P)&=& de-\tau^a \omega_a.
\end{eqnarray}

As expected for a non-degenerate bilinear form, all curvatures vanish, thus the conventional constraints are satisfied automatically.
In this situation the local gauge transformations are equivalent to the diffeomorphisms \cite{VanProeyen}. Note that we have a two dimensional foliation of the space subject to the constraint \eqref{motionRN1}.

Now, we are interested in expressing the spin connections $\omega_\mu$ and ${\omega_\mu}^a$ in terms of the geometric fields $e$ and $\tau_\mu$. 
For instance, the spin connection $\omega_\mu$ can be obtained from \eqref{motionRN1} by means of the following identity
\begin{equation}
R_{a\mu \nu}(H) {\tau_\rho}^a+R_{a \rho \mu}(H) {\tau_\nu}^a-R_{a \nu \rho}(H) {\tau_\mu}^a=0.
\end{equation}
Then, taking into account the relations \eqref{relations} on this last equation we get
\begin{equation}\label{spinconnection}
\omega_\mu=-\frac{1}{2}\epsilon_{ab}\left( \partial_{[\mu} {\tau_{\nu]}}^a \tau^{\nu b}-\partial_{[\mu}{\tau_{\nu]}}^b \tau^{\nu a}+ \partial_{[\lambda}{\tau_{\sigma]}}^c \tau^{\lambda a}\tau^{\sigma b}{\tau_\mu}^c \right),
\end{equation}
which coincides with that of \cite{Andringa:2012uz} in three dimensions.
On the other hand, considering the following identity
\begin{equation}
R_{\mu\nu}(P)e^\mu \tau^{\nu b}=0,
\end{equation}
we find that the spin connection ${\omega_{\mu}}^{a}$ cannot be completely determined, i.e., 
\begin{equation}
{\omega_{\mu}}^a={{\omega_\perp}_{\mu }}^a+{{\omega_\parallel}_{\mu }}^a
\end{equation} where the determined part is 
\begin{equation}\label{spinperp}
{{\omega_\perp}_{\mu }}^a= -\partial_{[\lambda}e_{\nu]}e^\lambda \tau^{\nu a} e_\mu. 
\end{equation} whic is perpendicular to $\tau^{\mu a}$.

 In a second order formulation, the undetermined part ${\omega_{\parallel \mu }}^a$ can be interpreted as a Lagrange multiplier, since the geometric constraint on the torsion $R^{a}(H)=0$ arises from the variation of \eqref{SGaction} with respect to this field. This situation is common in the context of second order NR theories for gravity, for example, the same situation happens for Carroll and Galilei gravity \citep{Bergshoeff:2017btm} and extended string Newton-Cartan gravity \cite{Bergshoeff:2018vfn}. 
In fact, the action \eqref{SGaction} in the second order formulation reads
\begin{equation} \label{ActionSO}
S[\tau,e]= \int \lbrace \gamma_1 \epsilon^{\mu \nu \rho}[\epsilon_{ab} {\tau_\mu}^a {{{\overset{\circ}{R}}}_{\nu \rho}}^b(M)-{{\omega_\parallel}_\mu}^a {\overset{\circ}{R}}_{ \nu \rho a}(H)-2\omega_\mu\partial_\nu e_\rho]+ \gamma_2 \epsilon^{\mu \nu \rho}\omega_\mu \partial_\nu \omega_\rho\rbrace  d^3 x,
\end{equation}
where,
\begin{eqnarray}
{{\overset{\circ}{R}}_{\mu \nu}}^a(M)&=& \partial_\mu {{\omega_\perp}_\nu}^a- \partial_\nu {{\omega_\perp}_\mu}^a-\epsilon^{ab}(\omega_\mu \omega_{\perp \nu b}-\omega_\nu \omega_{\perp \mu b}),\\
{\overset{\circ}{R}}_{\mu \nu a}(H)&=& \epsilon_{ab}(\partial_\mu {{\tau}_\nu}^b-\partial_\nu {{\tau}_\mu}^b)-(\omega_\mu \tau_{\nu a}-\omega_\nu \tau_{\mu a}),
\end{eqnarray}
with ${{\omega_\parallel}_{\mu }}^a$ and ${{\omega_\perp}_{\mu }}^a$ given in Eqs. \eqref{spinconnection} and \eqref{spinperp}, respectively.  In order to determine all the compoentes of the spin connection, as usual \cite{Andringa:2010it}, we add more generators  to the symmetry algebra. In fact, we can consider 
the stringy Galilei algebra with two non-central extensions a vector $Z_a$ and a pseudoscalar $Z$ \cite{Andringa:2012uz}\footnote{See the 
(\ref{ENH-algebra-Appendix}) in the Appendix when the dS (AdS) radius $\ell$ 
goes to infinity.}.
Then, considering these extra generators the most general invariant bilinear form for this algebra is 

\begin{equation}\label{PairingESG}
\langle H_a,M_b\rangle=\beta_1 \epsilon_{ab}, \quad \langle N,P \rangle=-\beta_2, \quad \langle N,N \rangle=\beta_3,
\end{equation}
\begin{equation}
 \quad \langle M_a,M_b \rangle=-\beta_4 \eta_{ab},\quad \langle Z,N \rangle =\beta_5.
\end{equation} 

 This bilinear form is degenerate for generic real values of $\beta_i$. This would imply that one that some of the curvatures associated to the gauge fields are not determined by the fields equations (as should be expected in a Chern-Simons theory), and some of the spin connections could be undermined. For more details, see the  example of  3d CS Maxwell
 gravity \cite{Aviles:2018jzw}.

\section{Stringy Newton-Hooke algebra} 

The $(2+1)$-Stringy Newton-Hooke $\mathfrak{snh}$ algebra is an In\"{o}n\"{u}-Wigner contraction of the $(2+1)$-dimensional dS (AdS) algebra.

The algebra is defined by the following non-vanishing commutators

\begin{eqnarray}\label{NH-algebra1}
\nonumber \left[ H_a , H_b\right] & =& \pm \frac{1}{\ell^2} \epsilon_{ab} N,  \quad \quad \left[N,H_a\right]  =  {\epsilon_a}^b H_b ,\\
\nonumber \left[H_a,M_b\right]  & = &   - \eta_{ab}P, \quad \quad \quad \left[N,M_a \right]  =  {\epsilon_a}^b M_b,\\
\left[P,H_a\right] & = & \pm \frac{1}{\ell^2} M_a,
\end{eqnarray}
where the upper sign refers to AdS and the lower sign to dS space-time and $\ell$ the dS (AdS) radius. These generators admit a bilinear form invariant and non-degenerate, however, again the spin connection ${\omega_{\mu}}^a$ cannot be completely determined from the field equations of stringy Newton-Hooke CS gravity.   

One interesting aspect  of the $\mathfrak{snh}$-algebra is that in $(2+1)$-dimension admits at least  two  non-central extensions \cite{Brugues:2006yd}. The NR limit of a string moving in a dS (AdS) background leads to a non-relativistic action which is invariant under the stringy Newton-Hooke symmetries with extensions.

Here we will obtain this algebra by applying the double extension procedure following the original works \cite{medina,Figueroa-OFarrill:1995opp} and \cite{Matulich:2019}. The procedure is explained
 in the Appendix. These non-central extensions can be obtained reducing to the 3-dimensional case of Ref. \cite{Brugues:2006yd} . 
The extended $\mathfrak{snh}$ algebra is completed adding to the algebra \eqref{NH-algebra1} the following commutators 
\begin{eqnarray}\label{Extended-part}
\nonumber && \left[P,M_a\right] = Z_a,  \quad \quad \quad  \quad \quad \nonumber \left[Z, H_a \right] = {\epsilon_a}^b Z_b, \quad \quad \left[N,Z_a \right] ={ \epsilon_a}^b Z_b ,\\ 
\nonumber &&  \left[M_a, M_b\right] = -\epsilon_{ab} Z, \quad \quad \left[H_a, Z_b\right] = \pm \frac{1}{\ell^2}\epsilon_{ab} Z . \\
\end{eqnarray}

Then, according the results of the Appendix the algebra in \eqref{NH-algebra1} and \eqref{Extended-part} admits the following invariant and non-degenerate bilinear form 
\begin{eqnarray}\label{NH-pairing}
\nonumber \langle H_a, H_b  \rangle & = &  \pm \frac{\alpha_2}{\ell^2} \eta_{ab}, \quad \quad \langle H_a, Z_b \rangle  =   \pm \frac{\alpha_3}{\ell^2} \eta_{ab}, \\
\nonumber \langle H_a, M_b \rangle & = &  \alpha_1 \epsilon_{ab}, \quad \quad\quad  \langle M_a, M_b \rangle = - \alpha_3 \eta_{ab}, \\
\nonumber \langle N, P \rangle & =  & -\alpha_1 , \quad \quad  \quad \quad \langle N, Z \rangle = \alpha_3 , \\
 \langle N, N \rangle & = & \alpha_2 , \quad \quad \quad \quad \quad \langle P , P \rangle = \pm \frac{\alpha_3}{\ell^2} .
\end{eqnarray}

A gauge-invariant gravity action invariant under the extended stringy Newton-Hooke algebra can be constructed using the one-form connection  
\begin{equation} \label{ConnectionNH}
A = \tau^a H_a + \omega^a M_a + \omega N + e P+ m^a Z_a + m Z, 
\end{equation}
whose components $\tau^a$, $\omega^a$, $\omega, e, m^a$ and $m$ are one-forms. The two-form curvatures associated to this gauge connection are given by
\begin{equation} \label{CurvatureNH}
R(A)=R^a(H_b)H_{a}+ R^a(M_b) M_{a}+R(N)N + R(P)P+ R^a(Z_b)Z_a + R(Z)Z,
\end{equation} 
where,
\begin{eqnarray}
R^a(H_b)&=& d\tau^a-\epsilon^{ab}\omega \tau_b ,\\
R^a(M_b)&=& d\omega^a-\epsilon^{ab}\omega \omega_b \pm \frac{1}{\ell^2} e \tau^a,\\
R(N)&=& d\omega \pm \frac{1}{2\ell^2} \epsilon_{ab} \tau^a \tau^b,\\
R(P)&=& de-\tau^a \omega_a, \\
R^a (Z_b) & = & dm^a - \epsilon^{ab} m \tau_b - \epsilon^{ab} \omega m_b +e \omega^a, \\
R(Z) & = & dm -\frac{1}{2} \epsilon_{ab} \omega^a \omega^b \pm \frac{1}{\ell^2} \epsilon_{ab}\tau^a m^b .
\end{eqnarray}

The dynamics of the gauge potential $A$ is described by the CS action

\begin{eqnarray} \label{NHaction}
\nonumber S[A]& = &
\int \left[   2\alpha_1 \left(  \epsilon_{ab} R^a (H_c) \omega^b - eR(N) \right) + \alpha_2 \left(   \omega R(N) \pm \frac{1}{\ell^2} \tau_a R^a (H_c) \mp \frac{1}{2\ell^2} \epsilon_{ab}\omega \tau^a \tau^b   \right) \right. \\
&&\left.+ \alpha_3 \left(  2m R(N) \pm \frac{1}{\ell^2} eR(P) - \omega_a R^a( M_b) \pm \frac{2}{\ell^2} m_a R^a (H_b)  \right) \right] .
\end{eqnarray}

This last action is quasi-invariant, i.e.,  up to boundary terms, under the action of the infinitesimal gauge transformations 
\begin{equation}\label{gaugetransformation}
\delta A_\mu = \partial_\mu \Lambda+[A_\mu,\Lambda], 
\end{equation}
with the local gauge parameter 
\begin{equation}
\Lambda= \lambda^a H_a +\Omega^a M_a + \Omega N + \lambda P + \Theta^a Z_a + \Theta Z.
\end{equation}

An analogue of the  $\alpha_1$ term was found in \cite{Obers:2016} for the particle case. Moreover, we see the explicit appearance of the gauge field $m^a$ in the $\alpha_3$ term in comparison to \cite{Bergshoeff:2018yvt}. The equations of motion are given by
\begin{eqnarray} \label{motionEq}
\delta_{\tau^a}S:\quad &&2  \alpha_1 R^a(M_b) \pm \frac{2\alpha_2}{\ell^2} R^a(H_b) \pm 2 \alpha_3 R^a (Z_b)= 0,\\ \label{motionRM}
\delta_{\omega^a} S: \quad && 2\alpha_1 R^a(H_b)-2\alpha_3 R^a(M_b)=0,\\ \label{motionRN}
\delta_{\omega} S: \quad && -2\alpha_1 R(P) + \alpha_2 R(N) + 2\alpha_3 R(Z) = 0,\\ \label{motionRH}
\delta_e S: \quad && 2 \alpha_1 R(N) \pm \frac{2\alpha_3}{\ell^2} R(P)= 0, \\
\delta_{m} S: \quad && 2\alpha_3 R(N) = 0 , \\
\delta_{m^a} S: \quad && \frac{2\alpha_3}{ \ell^2} R^a (H_b) =0.
\end{eqnarray}

From \eqref{gaugetransformation}, the gauge fields can be seen to transform as follows:
\begin{eqnarray} \label{NHtransformation}
\nonumber \delta {\tau_\mu}^a &=& \partial_\mu \lambda^a+ {\epsilon_b}^a \omega_\mu \lambda^b-{\epsilon_b}^a{\tau_{\mu}}^{b} \Omega,\\
\nonumber \delta {\omega_\mu}^a &=& \partial_\mu \Omega^a+ {\epsilon_b}^a \omega_\mu \Omega^b-{\epsilon_b}^a {\omega_\mu}^b \Omega \pm \frac{1}{\ell^2} e_\mu \lambda^a \mp \frac{1}{\ell^2} {\tau_\mu}^a \lambda,\\
\nonumber \delta \omega_{\mu} &=& \partial_\mu \Omega \pm \frac{1}{\ell^2} \epsilon_{ab}{\tau_\mu}^a \lambda^b,\\
\nonumber \delta e_\mu&=& \partial_\mu \lambda +{\omega_\mu}^a \lambda_a - {\tau_\mu}^a \Omega_a, \\
\nonumber \delta m_\mu & = & \partial_\mu \Theta \pm \frac{1}{\ell^2} \epsilon_{ab} {\tau_\mu}^a \Theta^b - \epsilon_{ab} {\omega_\mu}^a \Omega^b \pm \frac{1}{\ell^2} \epsilon_{ab} {m_\mu}^a \lambda^b, \\
\delta {m_\mu}^a & = & \partial_\mu \Theta^a + {\epsilon_a}^b m_\mu \lambda^b  + {\epsilon_b}^a \omega_\mu \Theta^b + e_\mu \Omega^a - {\epsilon_b}^a {m_\mu}^b \Omega - {\omega_\mu}^a \Omega.
\end{eqnarray}


Once again, we are interested in expressing the spin connections $\omega_\mu$ and ${\omega_\mu}^a$ in terms of the geometric fields 
${{\tau}_\mu}^a, e_\mu, m_\mu$ and ${{m}_\mu}^a$. 

In a similar way to the previous section the spin connection $\omega_\mu$ is obtained from
\begin{equation}
{R^a}_{\mu \nu}(H) {\tau_\rho}^a+{R^a}_{\rho \mu}(H) {\tau_\nu}^a-{R^a}_{\nu \rho}(H) {\tau_\mu}^a=0.
\end{equation}

Then, applying the inverse relations \eqref{relations},  we get
\begin{equation}
\omega_\mu=-\frac{1}{2}\epsilon_{ab}(\partial_{[\mu} {\tau_{\nu]}}^a\tau^{\nu b}-\partial_{[\mu}{\tau_{\nu]}}^b\tau^{\nu a}+\partial_{[\lambda}{\tau_{\sigma]}}^c\tau^{\lambda a}\tau^{\sigma b}{\tau_\mu}^c),
\end{equation} 
which was obtained previously in (\ref{spinconnection}). It is a straightforward check that the other spin connection ${{\omega}_\mu}^a$ can be determined completely as well. If we impose the relation
\begin{equation}
R_{\mu\nu}(P)e^\mu \tau^{\nu b}=0,
\end{equation}
and considering \eqref{relations} we find that
\begin{equation}\label{eqomega1}
e^\mu {\omega_{\mu}}^b = -2\tau^{\nu b} e^\mu \partial_{[\mu } e_{\nu ]}.
\end{equation}
Now, we can if we also impose
\begin{equation}
{R^a}_{\mu \nu} (Z_b)e^\nu \tau^{\mu c} = 0 ,
\end{equation}
yields,
\begin{equation}\label{eqomega2}
\tau^{\nu c} {\omega_{\nu}}^a = 2 \tau^{\mu c} e^\nu \left( \partial_{[ \mu} {m_{\nu]}}^a -\epsilon^{ab} \omega_{[\mu} m_{\nu] b}   \right) +\epsilon^{ac} e^\nu m_\nu.
\end{equation}
Combining equations \eqref{eqomega1} and \eqref{eqomega2} for the different projections and using the decompositions 
\begin{equation}
{\omega_{\mu}}^a={\tau_\mu}^b {\tau^{\nu}}_b {\omega_\nu}^a +e_\mu e^\nu {\omega_{\nu}}^a,
\end{equation}
we get
\begin{equation}
{\omega_\mu}^{a} = 2 {\tau_\mu}^{c}  {\tau^\lambda}_c e^\sigma \left( \partial_{[ \lambda} {m_{\sigma]}}^a -\epsilon^{ab} \omega_{[\lambda} m_{\sigma] b}   \right) + \epsilon^{ab} \tau_{\mu b}e ^\nu m_\nu -2e_\mu\tau^{\sigma a} e^\lambda\partial_{[\lambda } e_{\sigma ]} ,
\end{equation}
which coincides with that of \cite{Andringa:2012uz} in three dimensions.

As we see,  the equations used to determine these spin connections do not correspond to the ones obtained from variation with respect to $\omega$ and $\omega^a$. Thus, if we substitute these expressions in the first order action,  the second order Lagrangian will have a different dynamics, as in the ``exotic'' gravity or massive gravity \cite{Witten:1988},
for a general discussion about the equivalence of first and second order formalism see for example \cite{Bergshoeff:2018luo}.

\section{Conclusion and Outlook}

In this work we have constructed two NR stringy gravities. For the case of the  Stringy Galilei algebra without extensions we found a CS action which does not allow to express all the spin connections in terms of the dreibeins. These spin connections are Lagrange multipliers in a second order nonrelativistic theory.
Instead, if we consider the algebra with extensions, the bilinear form associated to this new algebra turns out  to be degenerate and therefore not all the curvatures vanish.

We have applied the same gauging procedure for the stringy Newton-Hooke algebra with two non-central extensions. In this case we can express
the spin connections in terms dreibeins and the gauge field associated to $Z_a$. The CS action contains all the gauge
fields including the gauge field $m_{\mu}$ associated to the scalar central charge $Z$ that was not present in the action for 
Newton-Hooke string \cite{Brugues:2006yd} and in the gravity action studied in \cite{Bergshoeff:2018vfn}.

We leave for future work the construction of a second order gravity action and the connection with possible relativistic gravity theories. It will be also interesting to see if using the Galilean free algebra \cite{galileanfree} in three dimensions one can construct 
a CS action as well.

\section*{Acknowledgements}
We are grateful to Axel Kleinschmidt, Jakob Palmkvist, Jakob Salzer and Jorge Zanelli for useful discussions. In particular, we are grateful to Jorge Zanelli for a careful reading of the paper.
DH and LA  are partially founded by Conicyt grants 21160649 and 21160827, respectively. The Centro de Estudios Cient\'ificos (CECs) is funded by the Chilean Government through the Centers of Excellence Base Financing Program of Conicyt. LA thanks the Departament de F\'isica Qu\`{a}ntica i Astrof\'isica of Universitat de Barcelona for hospitality. 
 JG has been supported in part by MINECO FPA2016-76005-C2-1-P and Consolider CPAN, and by the Spanish government (MINECO/FEDER) under project MDM-2014-0369 of ICCUB (Unidad de Excelencia Mar\'{i}a de Maeztu).

\appendix

\vspace{0.8cm}

{\Large {\bf Appendix}}
\section{Double Extensions}
A {\it double extension} $\mathfrak{d}$ 
 of a Lie algebra $\mathfrak{G}$ is an enlargement by means of one or more central or non-central extensions which has associated a non-degenerate invariant symmetric bilinear form $\Omega^{\mathfrak{d}}$ \cite{medina} \cite{Figueroa-OFarrill:1995opp}, (see also \cite{Matulich:2019}). 
 \\

\subsection{Procedure} 

\begin{itemize}

\item Firstly \footnote{We will follow the notation described in Ref. \cite{Matulich:2019}.}, suppose a Lie  subalgebra $\mathfrak{g} \in \mathfrak{G}$ equipped with a non-degenerate invariant metric $\Omega^{\mathfrak{g}}$ where its generators $G_i$ satisfy
\begin{eqnarray} \label{invariance1}
 [G_i, G_j ] =  {f_{ij}}^k G_k , \quad \quad  \big< G_i, G_j \big>_{\mathfrak{g}} = \Omega^{\mathfrak{g}}_{ij} , \quad \quad {f_{ij}}^k \Omega^{\mathfrak{g}}_{kl}  	+{f_{il}}^k \Omega^{\mathfrak{g}}_{kj} = 0.
\end{eqnarray}

Here $\Omega^{\mathfrak{g}}_{ij}$ denotes the symmetric matrix form associated to the non-degenerate invariant metric $\Omega^{\mathfrak{g}}$ and ${f_{ij}}^k$ the structure constants of $\mathfrak{g}$.  
\item Secondly, let us introduce a new Lie subalgebra $\mathfrak{h}$ which act on $\mathfrak{g}$ via 
\begin{equation} \label{invariance2}
[H_\alpha  , H_\beta] = {f_{\alpha \beta}}^\gamma H_\gamma,\quad [H_\alpha  , G_i] = {f_{\alpha i}}^j G_j, \quad {f_{\alpha i}}^k \Omega^{\mathfrak{g}}_{kj} + {f_{\alpha j}}^k \Omega^{\mathfrak{g}}_{ki} =0,
\end{equation}

where $H_\alpha \in \mathfrak{h}$ and ${f_{\alpha \beta}}^\gamma$ denote the structure constants of $\mathfrak{h}$. We say that $\mathfrak{g}$ is a representation of $\mathfrak{h}$. 
Notice that it is not necessary to impose constraints on the the invariant symmetric bilinear form of $\mathfrak{h}$, say $h_{\alpha \beta}$, i.e., it  could eventually to be degenerate or zero. 
\item Finally, let us introduce the dual of $\mathfrak{h}$, to call $\mathfrak{h}^{\star}$ with its generators $H^{\star \alpha}$.\\ 

We call a {\it double extension} of $\mathfrak{G}$ to the Lie algebra $\mathfrak{d}=D(\mathfrak{g}, \mathfrak{h})$ defined on the vector space $\mathfrak{g} \oplus \mathfrak{h} \oplus \mathfrak{h}^{\star}$ as a semidirect sum	 with the following commutation relations
\begin{equation} \label{importanteq}
\left[G_i, G_j \right] = {f_{ij}}^k G_k + {f_{\alpha i}}^k \Omega^{\mathfrak{g}}_{kj} H^{\star \alpha},
\end{equation}
\begin{eqnarray}
\left[ H_\alpha , H^{\star \beta} \right]  & = &  - {f_{\alpha \gamma}}^{\beta} H^{\star \gamma} , \\
\left[ H^{\star \alpha} , G_j \right]  & = & 0, \\
 \left[ H^{\star \alpha} , H^{\star \beta} \right]  &  =& 0 .
\end{eqnarray}

\end{itemize}
As we see in Eq.\eqref{importanteq} the action of $\mathfrak{h} $ on $\mathfrak{g}$ fixes the structure constants ${f_{\alpha i}}^k$. The Lie algebra $\mathfrak{d}$ admits a non-degenerate metric written in a matrix form as follows
\begin{equation}
\Omega^{\mathfrak{d}}_{ab} = \left( \begin{array}{ccc}
\Omega^{\mathfrak{g}}_{ij} & 0 & 0 \\
0 & h_{\alpha \beta} & {\delta_\alpha}^\beta \\
0 & {\delta^\alpha}_\beta & 0
\end{array}
 \right).
\end{equation}
As we mentioned, the metric associated to the algebra $\mathfrak{h}$ can be degenerate or zero, so that it does not affect the degeneracy of $\Omega^{\mathfrak{d}}$. \\
In the following, we will see the applications of this method to the two algebras of interest in this paper, namely the stringy Galilei and Newton-Hooke algebras.

\subsection{Stringy Galilei algebra}

For the case of Stringy Galilei algebra, we first choose
\begin{eqnarray}\label{choosestringygalilei}
\nonumber  \mathfrak{g}_1 & =  & \{  P , M_a  \} , \\ 
\mathfrak{h}_1 &  = & \{ N, H_a \} ,
\end{eqnarray}
with $\mathfrak{g}_1$ being an abelian subalgebra, and the commutation relations of $\mathfrak{h}_1$ are given by

\begin{eqnarray}
\left[ H_a , H_b\right] & =& 0,  \quad \quad \left[N,H_a\right]  =  {\epsilon_a}^b H_b .
\end{eqnarray}

The representation of $\mathfrak{g}_1$ with respect to $\mathfrak{h}_1$ is
\begin{eqnarray}\label{NH-algebra}
\left[N,M_a \right]  =  {\epsilon_a}^b M_b, \quad \left[H_a,M_b\right]   =    - \eta_{ab}P,
\quad \left[P,H_a\right]  =  0.\quad  \left[P,N\right]  =  0 .
\end{eqnarray}

Despite that the invariance of $\big< , \big>_{\mathfrak{g}_1}$ with respect to $\mathfrak{g}_1$ is satisfied, it is not possible to fulfil with respect to $\mathfrak{h}_1$, i.e.,  the invariance \eqref{invariance2}. In fact, 
\begin{eqnarray}
 \big< [ H_a ,P], M_b \big>_{\mathfrak{g}_1}  +  \big< P,[H_a ,M_b]\big>_{\mathfrak{g}_1} =
 0-\delta_{ab}\big<  P,  P\big>_{\mathfrak{g}_1}\neq 0.
\end{eqnarray}

This last equation is in contradiction with the invariance $\big< , \big>_{\mathfrak{g}_1}$ with respect to $\mathfrak{g}_1$. Then, we cannot continue with the rest of the procedure. Analogously, for the chooses: $\mathfrak{g}_1=\lbrace H_a,P \rbrace$ with $\mathfrak{h}_1=\lbrace N, M_a \rbrace$, the metric $\big< , \big >_{\mathfrak{g}_1}$ is not invariant because $M_b \big< H_a, P 	\big> = \eta_{ab} \big< P , P \big>_{\mathfrak{g}_1} \neq 0$; and $\mathfrak{g}_1=\lbrace H_a, M_a,P \rbrace$ with $\mathfrak{h}_1=\{ N \}$ we have that the metric of $\mathfrak{g}_1$ is not invariant because $H_a \big< P, M_b \big>_{\mathfrak{g}_1} =-\eta_{ba} \big< P, P \big>_{\mathfrak{g}_1} \neq 0$. Again, we cannot continue with the procedure. Notice that there are no other cases in which $\mathfrak{h}_1$ act on $(\mathfrak{g}_1,\Omega^{\mathfrak{g}_1})$ via antisymmetric derivations. \\

Therefore, for the case of Stringy Galilei algebra is not possible to perform the double extension procedure starting from this algebra. However, we have a non-degenerate-invariant metric in $\mathfrak{g}_1 $ with the choose in \eqref{choosestringygalilei}. \\

\subsection{Stringy Newton-Hooke algebra}

	Now, we will apply this procedure for the $\mathfrak{snh}$ algebra without extensions \cite{Brugues:2006yd} 
	\begin{eqnarray}\label{ENH-algebra-Appendix}
\nonumber \left[ H_a , H_b\right] & =& \pm \frac{1}{\ell^2} \epsilon_{ab} N, \\
\nonumber \left[N,H_a\right] & = & {\epsilon_a}^b H_b, \\
\nonumber \left[N,M_a \right] & = & {\epsilon_a}^b M_b, \\
\nonumber \left[H_a,M_b\right] & = &  - \eta_{ab}P, \\
\left[P,H_a\right] & = & \pm \frac{1}{\ell^2} M_a.
\end{eqnarray}

In this case we choose the following sub-algebras 
 	 \begin{eqnarray}
\mathfrak{g}_2 & =  & \{  \widetilde P, M_a  \} ,\\
\mathfrak{h}_2 &  = & \{ N, \widetilde H_a \} ,
\end{eqnarray}

where we have defined the generators $\widetilde P = \ell P$ and $\widetilde H_a = \ell H_a$. Notice that there are no more choices for the construction of these sub-algebras satisfying the \eqref{invariance1}, \eqref{invariance2} and \eqref{importanteq} constraints. 
It can be shown that the non-degenerate-invariant metric in $\mathfrak{g}_2 $ is invariant and non-degenerate with respect to $\mathfrak{g}_2$ and $\mathfrak{h}_2$, and is given by
\begin{eqnarray}
\big< \widetilde P , \widetilde P \big>_{\mathfrak{g}_2} =1,\quad \big< M_a , M_b \big>_{\mathfrak{g}_2} =\delta_{ab},\quad \big< \widetilde P, M_a \big>_{\mathfrak{g}_2}=0 .
\end{eqnarray} 

The non-vanishing commutators of the extended algebra yield
\begin{eqnarray}
\left[  \widetilde{P}, M_a \right]&  = &  \mp H^{\star \widetilde{H}}_{a}, \\
\left[M_a,M_b \right] & = & - \epsilon_{ab} H^{\star N}, \\
\left[ N,Z_a \right] & = & \mp {\epsilon_a}^b H_{a}^{\star \widetilde{H}}, \\ 
\left[ \widetilde{H}_a, Z \right] &  = & \pm {\epsilon_a}^b H^{\star \widetilde{H}}, \\
\left[ \widetilde{H}_a, Z_b \right] & = & \pm \epsilon_{ab} H^{\star N} . 
\end{eqnarray}

Defining the generators $Z_a:=	\mp  H^{\star \widetilde H}_{a} $ and $Z:=H^{\star N} $ and reinserting the cosmological constant $\Lambda=  \displaystyle \frac{1}{\ell^2}$, we get

\begin{eqnarray}\label{ENH-algebra-Appendix}
\nonumber \left[ H_a , H_b\right] & =& \pm \frac{1}{\ell^2} \epsilon_{ab} N,  \quad \quad \quad \left[P,M_a\right] = Z_a, \\
\nonumber \left[N,H_a\right] & = & {\epsilon_a}^b H_b ,  \quad \quad \quad  \quad \left[Z, H_a \right] = {\epsilon_a}^b Z_b, \\
\nonumber \left[N,M_a \right] & = & {\epsilon_a}^b M_b,  \quad  \quad \quad \quad \left[N,Z_a \right] ={ \epsilon_a}^b Z_b, \\
\nonumber \left[H_a,M_b\right] & = &  - \eta_{ab}P,  \quad \quad \quad   \left[M_a, M_b\right] = -\epsilon_{ab} Z, \\
\left[P,H_a\right] & = & \pm \frac{1}{\ell^2} M_a,  \quad \quad \quad \left[H_a, Z_b\right] = \pm \frac{1}{\ell^2} \epsilon_{ab} Z,
\end{eqnarray}
with the following non-degenerate invariant bilinear form 
\begin{eqnarray}\label{NH-pairing}
\nonumber \langle H_a, H_b  \rangle & = &  \pm \frac{\alpha_2}{\ell^2} \eta_{ab}, \quad \quad \langle H_a, Z_b \rangle  =   \pm \frac{\alpha_3}{\ell^2} \eta_{ab}, \\
\nonumber \langle H_a, M_b \rangle & = &  \alpha_1 \epsilon_{ab}, \quad \quad\quad  \langle M_a, M_b \rangle = - \alpha_3 \eta_{ab}, \\
\nonumber \langle N, P \rangle & =  & -\alpha_1 , \quad \quad  \quad \quad \langle N, Z \rangle = \alpha_3 , \\
 \langle N, N \rangle & = & \alpha_2 , \quad \quad \quad \quad \quad \langle P , P \rangle = \pm \frac{\alpha_3}{\ell^2} ,
\end{eqnarray}
with $\alpha_i$ arbitrary real constants and $\alpha_3 \neq 0$ for the matrix to be non-degenerate.
 We call to \eqref{ENH-algebra-Appendix} the (2+1)-Newton-Hooke algebra with extensions.  Of course, if we 
 further take the limit $\ell \rightarrow \infty$ we obtain the stringy Galilean algebra with extensions. However, 
 in this limit the symmetric bilinear form turns out to be degenerated. Therefore, 
 we can observe the important role of the cosmological constant at the moment to write down a well defined Chern-Simons action. \\

\newpage 

	


\end{document}